\documentclass{article}
\usepackage[pdftex]{graphicx}
\usepackage{arxiv}
\usepackage{algorithm}
\usepackage{algorithmic}
\usepackage{graphicx}    
\usepackage{natbib}       
\usepackage{color}
\usepackage{url}
\usepackage{amsthm}
\newtheorem{assupm}{Assumption}
\newtheorem{prb}{Problem}

\newtheorem{fact}{Fact}
\newtheorem{lemmalemma}{Lemma}

\newtheorem{thmthm}{Theorem}

\newtheorem{rem}{Remark}
\theoremstyle{remark} 
\newtheorem*{pf}{Proof}
\usepackage{framed}
\usepackage{fancybox}
\usepackage{framed}
\usepackage{amsmath}
\usepackage{ascmac}
\usepackage{hyperref}       
\usepackage[utf8]{inputenc} 
\usepackage[T1]{fontenc}    
\usepackage{url}            
\usepackage{booktabs}       
\usepackage{amsfonts}       
\usepackage{nicefrac}       
\usepackage{microtype}      
\usepackage{cleveref}       
\usepackage{lipsum}         
\usepackage{doi}
\usepackage[normalem]{ulem}
\usepackage{XCharter}
\usepackage[T1]{fontenc}
\usepackage{dsfont}

\title{\scalebox{0.8}{arXiv}\\Cloud-mediated Self-triggered Synchronization of\\ a General Linear Multi-agent System\\ over a Directed Graph} 
\newif\ifuniqueAffiliation
\uniqueAffiliationtrue

\ifuniqueAffiliation 
\author{ \href{https://orcid.org/0009-0008-2906-2711}{\includegraphics[scale=0.02]{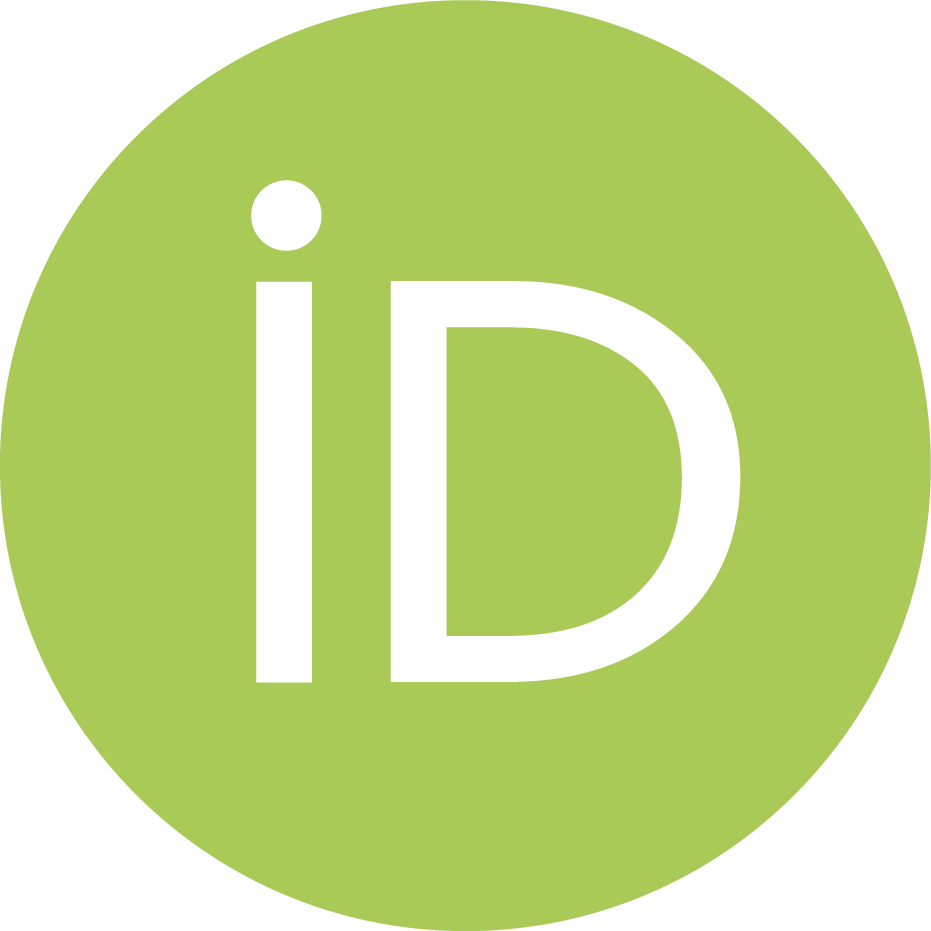}\hspace{1mm}Takumi Namba}\\	
	 (ORCID: 0009-0008-2906-2711)\\
Dept. of Electrical and Electronic Engineering\\
	Ritsumeikan University\\
	1-1-1 Noji-Higashi, Kusatsu, Shiga, Japan \\
	\texttt{re0075hf@ed.ritsumei.ac.jp} \\
	\And
	\href{https://orcid.org/0000-0002-1272-9486}{\includegraphics[scale=0.02]{orcid.eps}\hspace{1mm} Kiyotsugu Takaba} \\(ORCID: 0000-0002-1272-9486)\\
Dept. of Electrical and Electronic Engineering\\
	Ritsumeikan University\\
	1-1-1 Noji-Higashi, Kusatsu, Shiga, Japan \\
	\texttt{ktakaba@fc.ritsumei.ac.jp}\\
	}
\else
\fi
\renewcommand{\shorttitle}{Namba and Takaba (2023, \textit{arXiv})}
\begin{document}
\maketitle
\begin{abstract}                       
This paper proposes a self-triggered synchronization control method of a general high-order linear time-invariant multi-agent system through a cloud repository. In the cloud-mediated self-triggered control, each agent asynchronously accesses the cloud repository to get past information on its neighboring agents. Then, the agent predicts future behaviors of its neighbors as well as of its own, and locally determines its next access time to the cloud repository. In the case of a general high-order linear agent dynamics, each agent has to estimate exponential evolution of its trajectory characterized by eigenvalues of a system matrix, which is different from single/double integrator or first-order linear agents. Our proposed method deals with exponential behaviors of the agents by tightly evaluating the bounds on matrix exponentials.  Based on these bound, we design the self-triggered controller through a cloud which achieves bounded state synchronization of the closed-loop system without exhibiting any Zeno behaviors. The effectiveness of the proposed method is demonstrated through the numerical simulation.
\end{abstract}
\section{Introduction}
Recently, coordination of a multi-agent system has been widely studied. In a multi-agent system, each agent autonomously acts through interaction with its neighboring agents, and the whole system achieves various cooperative tasks such as consensus and synchronization \citep{Li,Saber,Scardovi,Takaba,Trentelman}. Most of the existing works assume that each agent can simultaneously and continuously exchange local information with its neighboring agents. In view of practical situations, this is not realistic because of high energy consumption and communicational burden.

To tackle the above difficulties by reducing communication frequencies, \emph{event-triggered} control and \emph{self-triggered} control have been proposed recently in the literature \citep{Almeida,Dimarogonas10,Ding,Nowzari-survey,Yang,Zhu}. In the event-triggered control strategy, each agent updates its control input signal only when a prescribed triggering condition is satisfied. Moreover, a self-triggered controller computes next triggering times based on a prediction of future behaviors of the neighboring agents (see \cite{Dimarogonas10}, \cite{Ding} and the references therein). For example, \citet{Zhu} and \citet{Almeida} proposed event-triggered and self-triggered synchronization control strategies for linear multi-agent systems, respectively.  Most of the event-triggered controllers employed a peer-to-peer (P2P) communication scheme between agents. 

As another research direction, there have recently been reported several studies on self-triggered coordination through a \emph{cloud repository}, which does not require the instantaneous inter-agent P2P communication. In the control schemes using a cloud repository (hereafter, we sometimes call it ``\emph{a cloud}'' simply), each agent asynchronously accesses the cloud to get past information on its neighboring agents.  Then, the agent predicts the future behaviors of its neighboring agents as well as of its own, and locally determines its next access time\footnote{In this paper, we define the terms ``\emph{triggering time}'' and ``\emph{access time}'' as follows. The term ``triggering time'' means the time instant at which the control inputs are updated in the conventional event-triggered/self-triggered control schemes. On the other hand, the term ``access time'' is basically the time instant at which some agents access to the cloud repository in the cloud-mediated scheme. It is assumed that, at the same time of the access to the cloud, the control input signal to the corresponding agent is updated.} to the cloud. Actually, self-triggered control through a cloud is especially effective in the situation where the instantaneous communication between the agents is not possible. As mentioned in the literature, a typical example of such control systems is autonomous underwater vehicles (AUVs). The AUVs cannot communicate with each other due to limited ability of their IoT devices. Another benefit of the cloud-mediated control is that each agent does not have to be listening (keeping communication channels open) to the triggering of the neighboring agents. In the standard self-triggered control, each agent requires the neighbors' states at its triggering time to compute the consensus/synchronization input. This means that each agent requests the neighbors to send their current states. In practical situations, this means that each agent always has to wait for the triggering of the neighboring agents, and it is not desirable from the energy conservation point of view.

There have been several works on the cloud-mediated control for agents with very simple dynamics \cite{Adaldo15,Adaldo16,Adaldo17,Adaldo18,Bowman1,Nowzari16,Namba}. For the agents with single integrator dynamics, self-triggered consensus control methods via a cloud were proposed by \citet{Nowzari16} and \citet{Bowman1}. Adaldo~\textit{et~al.}~\cite{Adaldo15,Adaldo16} also studies the cloud-supported consensus and tracking control for the single integrators in the presence of disturbances. \citet{Adaldo17} also extended the above results to the cloud-supported formation control of the agents with double integrator dynamics. To the best of the author's knowledge, there have not been reported any results on the general linear agent case except for our preliminary work on the \emph{first-order} agents over the \emph{undirected} graph \citep{Namba}.

In this paper, we will consider a self-triggered synchronization control method for a  \emph{general high-order linear time-invariant (LTI)} multi-agent system through the cloud. One of the major difficulties in the control scheme using the cloud is that each agent cannot instantaneously communicate with its neighboring agents at every triggering time, and thus each agent has to accurately estimate current and future behaviors of its neighbors' states. In the case of a high-order \emph{LTI} agent dynamics, it is necessary to handle the exponential evolution of its trajectory characterized by the eigenvalues of a system matrix in contrast to the aforementioned single/double integrators and first-order linear cases. As one of our contributions through this work, we design a triggering function by tightly evaluating the bound on matrix exponentials. We will prove that the proposed method achieves the bounded state synchronization of the closed-loop system without exhibiting any Zeno behaviors. We will also characterize the theoretical lower bounds of access intervals based on the bound on the matrix exponential. It should be noted that these insights have not ever appeared in any previous works for integrator/double integrator cases as well as our preliminary result \cite{Namba}. 

\paragraph*{Notational Conventions: }We denote the $N$ dimensional all-ones vector by $\mathds{1}_N:=[1,\ldots,1]^{\sf T}$. $I_{N}$ is the identity matrix with the dimension $N$. Closed and open intervals are described by $[a,b]$ and $]a,b[$, respectively. (We use the similar notations for half-open intervals). $M\succ 0$ means that $M$ is positive definite, and $M\succeq 0$ means that $M$ is positive semidefinite. $X\otimes Y$ denotes the Kronecker product of $X$ and $Y$.

\section{Problem Formulation}
\subsection{Agent Model}
Consider $N$ LTI agents whose dynamics is given by the stabilizable state equation
$
\dot{x}_{i}(t)=Ax_{i}(t)+Bu_{i}(t),~i=1,\ldots,N,
$
where $x_{i}(t) \in\mathds{R}^n$ is the state, $u_{i}(t)\in\mathds{R}^m$ is the input, and $A\in\mathds{R}^{n\times n}$ and $B\in\mathds{R}^{n\times m}$ are constant matrices. We denote the collection of all agents' states and inputs by 
$
x(t):=
[
x_{1}^{\sf T},\ldots,x_{N}^{\sf T}
]^{\sf T}\in\mathds{R}^{Nn}, u(t):=
[u_{1}^{\sf T},\ldots,u_{N}^{\sf T}
]^{\sf T}\in\mathds{R}^{Nm}
$
respectively. 
\subsection{Communication Model}
\subsubsection{Cloud Repository}
In this sub-section, we introduce the model of a cloud repository due to \cite{Nowzari16, Adaldo16}.
As described in the previous section, each agent communicates with its neighbors through the cloud which only stores information of all the agents and does not execute any computation. We assume that all the operations of access to the cloud, i.e., connections, downloading/uploading, disconnections are instantaneous. In other words, there are no delays in the communication between the agents and the cloud.

The cloud stores the information described in Table \ref{tab:cloudstruc}.
We denote the latest access time of the agent $i$ by $t_{i}^{l_{i}(t)}$ at the current time $t$, where $l_{i}(t)$$\in\mathds{Z}_{\geq 0}$ indicates how many times the agent $i$ connected to the cloud before and at the time $t$. It may be noted that $l_{i}$ is viewed as the access counter. We often write $t_{i}^{l_{i}}$ by dropping the current time for simplicity. We also introduce the set $\mathds{A}_{i}:=\{l_{i}\} \subseteq \mathds{Z}_{\geq 0}$ which denotes the sequence of the number of the accesses.
\begin{table}[htb]
\begin{center}
\caption{Information stored in the cloud at $t$}
\label{tab:cloudstruc}
\begin{tabular}{c | c c c c}
\hline
\textbf{Agents} & $\mathbf{1}$ & $\mathbf{2}$ & $\cdots$ & $\boldsymbol{N}$ \\ \hline 
the last access time& $t_1^{l_{1}}$ & $t_{2}^{l_{2}}$ & $\cdots$ & $t_{N}^{l_{N}}$ \\ 
the last access state & $x_{1}(t_{1}^{l_{1}})$ & $x_{2}(t_{2}^{l_{2}})$ & $\cdots$ & $x_{N}(t_{N}^{l_{N}})$ \\ 
the input & $u_{1}(t_{1}^{l_{1}})$ & $u_{2}(t_{2}^{l_{2}})$ & $\cdots$ & $u_{N}(t_{N}^{l_{N}})$ \\
the next access time & $t_{1}^{l_{1}+1}$ & $t_{2}^{l_{2}+1}$ & $\cdots$ & $t_{N}^{l_{N}+1}$\\
\hline
\end{tabular}
\end{center}
\end{table}

Although the situation is possible that each agent can obtain all information stored in the cloud, we assume that each agent can access only information of the preassigned agents to keep the privacy of all agents. We adopt a mathematical graph to describe such communication model as in the ordinary multi-agent systems.
\subsubsection{Accessibility Graph}
Throughout this paper, it is assumed that each agent can only access the information of  predetermined agents among the information stored in the cloud repository. 

We employ a directed graph to describe which agents can be accessed by each agent. A mathematical graph has been widely used for describing communications between agents in conventional multi-agent systems (see e.g.~\cite{Bullo}). In this section, we briefly review the fundamentals of the algebraic graph theory which will be useful in this paper.

The accessibility between agents is modeled by a \emph{directed graph}, which is defined as a pair $\mathcal{G}=(\mathcal{V}, \mathcal{E})$, where $\mathcal{V}:=\{1,2,\ldots,N\}$ is a vertex set and $\mathcal{E}\subseteq\mathcal{V}\times \mathcal{V}$ is an edge set. Each vertex represents an agent and an edge $(j,i)\in\mathcal{E}$ means that the vertex $i$ can receive the information from the vertex $j$. 

\begin{figure}[htb]
\begin{center}
\includegraphics[width=6cm]{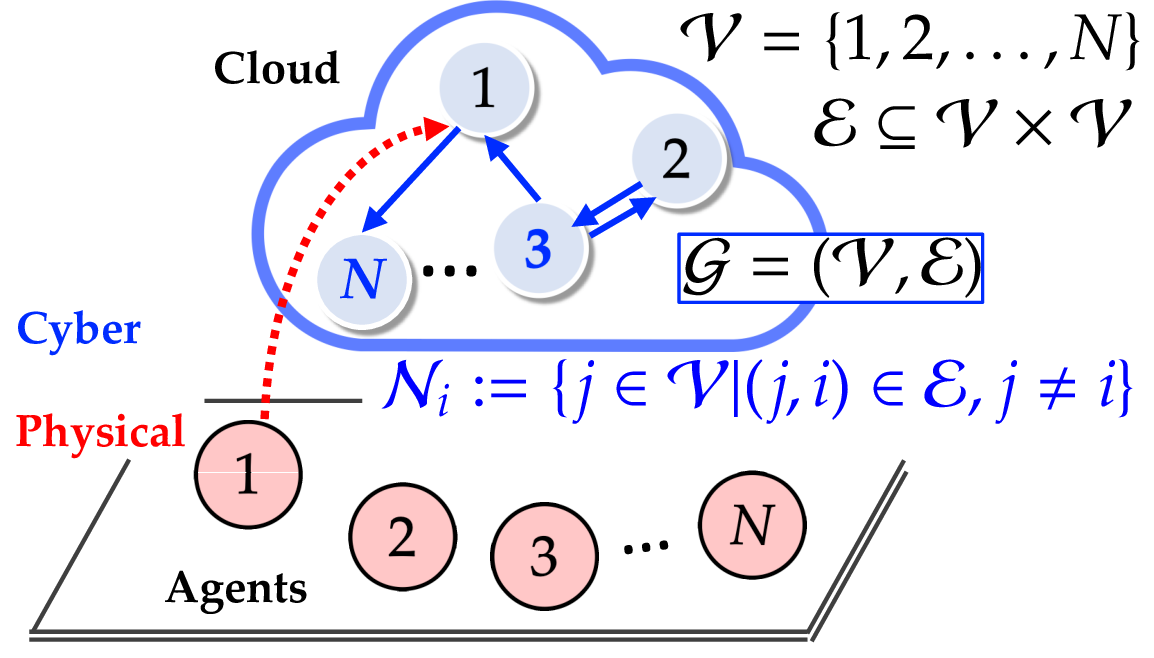}
\end{center}
\vspace{-1em}
\caption{Communication model}
\label{fig:communicationmodel}
\end{figure}

The overall communication model is depicted in Fig.~\ref{fig:communicationmodel}.
\begin{assupm}\mbox{}\\
\label{assupm:graph}
\vspace{-1em}
\begin{enumerate}
\renewcommand{\labelenumi}{(\roman{enumi})}
\item{The graph $\mathcal{G}$ has a directed spanning-tree.}
\item{The topology of the graph $\mathcal{G}$ is time-invariant.}
\end{enumerate}
\end{assupm}
We define the neighbor set of the agent $i$ by 
$\mathcal{N}_{i}:=\{j\in\mathcal{V}~|~(j,i)\in\mathcal{E},~j\neq i\}.$
We also define the graph Laplacian by 
$
L=[\ell_{ij}]\in\mathds{R}^{N\times N}$, where $\ell_{ij}=$
$|\mathcal{N}_{i}|$ if $i=j$, $-1$ if $(j,i)\in\mathcal{E}$, and otherwise $0$.
As is well known, the following fact holds true under Assumption \ref{assupm:graph} \citep{Bullo}. 
\begin{fact}\mbox{}\\
\label{fact1}
\vspace{-1em}
\begin{enumerate}
\renewcommand{\labelenumi}{(\roman{enumi})}
\item{$\mathrm{rank}L=N-1$, namely, $L$ has only one zero eigenvalue.}
\item{$\mathds{1}_N$ is the right eigenvector of the zero eigenvalue, namely, $L\mathds{1}_N=0$.}
\end{enumerate}
\end{fact}
For later discussion, let $\phi^{\sf T}$$\in\mathds{R}^{N}$ be the left eigenvector of the zero eigenvalue for $L$ such that $\phi^{\sf T}\mathds{1}_N=1$. It is seen from Gershgorin's disk theorem and the definition of $L$ that all the nonzero eigenvalues of $L$ have positive real parts. Hence, under Assumption \ref{assupm:graph} we denote the eigenvalues of $L$ with $\lambda_{1},\lambda_{2},\ldots,\lambda_{N}$ in the ascending order with respect to their real parts $
\lambda_{1}=0<\mathrm{Re}(\lambda_{2})\leq \mathrm{Re}(\lambda_{3})\leq\cdots\leq\mathrm{Re}(\lambda_{N}).$

Let us introduce the similarity transformation with a nonsingular matrix $U$ in the form of $U:=[\mathds{1}_{N} ~ X_{1}]\in$$\mathds{R}^{N\times N}$, $X_{1}\in\mathds{R}^{N\times(N-1)}$ which block-diagonalizes $L$ as $U^{-1}LU=
\mbox{diag}(
0, \check{L})$. The eigenvalues of $\check{L}\in$$\mathds{R}^{(N-1)\times(N-1)}$ are equal to $\lambda_{2},\ldots,\lambda_{N}$ because the similarity transformation preserves the eigenvalues of the original matrix. Note also that the first row of $U^{-1}$ is equal to $\phi^{\sf T}$.

\subsection{Problem Description}
We wish to solve the following \emph{bounded} synchronization problem. To formulate our synchronization problem, we define the synchronization errors as
\begin{align}
\label{eq:def-delta}
\delta_{i}(t)&:=x_{i}-\alpha\color{black},\\
\label{eq:def-alpha}
\alpha(t)&:=({\phi}^{\sf T}\otimes I_{n})x(t).
\end{align}
Note that $\delta(t)$ can be expressed as $\delta(t)=x(t)-\mathds{1}_N\otimes\alpha$ where $\delta(t):=[\delta_{1}^{\sf T},\ldots,\delta_{N}^{\sf T}]^{\sf T}\in\mathds{R}^{Nn}$.
\begin{prb}
For given constants $\eta_{0}>0$ and $\epsilon>0$, design a self-triggered control law through the cloud which achieves the bounded synchronization with the tolerance~$\epsilon$:
\begin{equation}
\color{black}
\label{eq:def-boundedsynchronization}
\|\delta(0)\|<\color{black}\eta_{0}\Rightarrow 
\lim_{t\rightarrow \infty} \sup \|\delta(t)\|\leq\epsilon.
\end{equation}
\end{prb}
\begin{rem}
\label{rem:uub}
According to \citet{Khalil}, the error dynamics of the closed-loop system is uniformly ultimately bounded (UUB) if there exist positive constants $b^{\natural}$ and $c^{\natural}$, for every $d^{\natural} \in \left]0,c^{\natural}\right[$ such that
, there is a positive constant $T^{\natural}=T(d^{\natural})$ satisfying 
\begin{align}
\label{eq:def-uub}
\|\delta(0)\|<d^{\natural} \Rightarrow \|\delta(t)\| \leq b^{\natural}~~\forall t\geq T^{\natural}.
\end{align}
As is obvious from the above problem statement, our self-triggered control law will achieve the uniformly ultimate boundedness of the synchronization error dynamics (take $b^{\natural}$ and $c^{\natural}$ such that $b^{\natural}\geq\epsilon$ and $c^{\natural}\leq \eta_{0}$).
\end{rem}
\section{Cloud-mediated Self-triggered Synchronization}
In this section, we will design a self-triggered synchronizing controller through the cloud.
Specifically, we will derive a triggering rule (access rule) and show that the proposed method 
does not exhibit Zeno behavior. Finally, we will conclude that the proposed controller 
achieves the bounded synchronization, and the closed-loops system is UUB.

\subsection{Controller Design}
In this sub-section, we design a self-triggered synchronizing controller following the line of \cite{Adaldo17,Adaldo18} and \cite{Almeida}.

Let us consider the \emph{fictitious} relative state feedback controller with the exact states $x_{i}(t)$ and $x_{j}(t)$, namely, $z_{i}(t)=F\sum_{j\in\mathcal{N}_{i}}(x_{j}(t)-x_{i}(t))$. As the agent $i$'s control input, we employ the relative state feedback controller under the zeroth-order hold (ZOH):
\begin{equation}
\label{eq:ut}
u_{i}(t)=F\sum_{j\in\mathcal{N}_{i}}(\hat{x}_{j}(t_{i}^{l_{i}})-x_{i}(t_{i}^{l_{i}})),~t\in[t_{i}^{l_{i}},t_{i}^{l_{i}+1}[,
\end{equation}
where $F\in\mathds{R}^{m\times n}$ is the gain to be designed, and $\hat{x}_{j}$ denotes the prediction of $x_{j}$ by the agent $i$, respectively. It should be noted that the agent $i$  cannot access $x_{j}(t_{i}^{l_{i}})$ but $x_{j}(t_{j}^{l_{j}})$. The agent $i$ predicts the neighbors' states $x_{j}$ by
\begin{equation}
\hat{x}_{j}(t)=
\begin{cases}
\mathrm{e}^{A(t-t_{j}^{l_{j}})}x_{j}(t_{j}^{l_{j}})+\int_{t_{j}^{l_{j}}}^{t}\mathrm{e}^{A(t-\tau)}Bu_{j}(t_{j}^{l_{j}})\mathrm{d}\tau & t\in[t_{j}^{l_{j}},t_{j}^{l_{j}+1(t_{i}^{l_{i}})}],\\
\mathrm{e}^{A(t-t_{j}^{l_{j}+1})}\hat{x}_{j}(t_{j}^{l_{j}+1(t_{i}^{l_{i}})}) & t>t_{j}^{l_{j}+1(t_{i}^{l_{i}})}.
\end{cases}
\label{eq:predictionsofxj}
\end{equation}
\normalsize
Recall that $t_{j}^{l_{j}+1(t_{i}^{l_{i}})}$ denotes the next access time of the agent $j$ at time $t_{i}^{l_{i}}$. It should be noted that the agent $i$ can compute the exact value of ${x}_{j}(t_{i}^{l_{i}})$, and hence we can replace $\hat{x}_{j}(t_{i}^{l_{i}})$ by $x_{j}(t_{i}^{l_{i}})$ in \eqref{eq:ut}. The input $u_{i}(t)$ is kept constant at $t\in[t_{i}^{l_{i}},t_{i}^{l_{i}+1}[$. Moreover, the second equation in \eqref{eq:predictionsofxj} is not used for the computation of the control input \eqref{eq:ut} at $t=t_{i}^{l_{i}}$ but for the self-triggering rule.
\begin{rem}
Different from the standard event-triggered and self-triggered control techniques, at the triggering time $t_{i}^{l_i}$, each agent $i$ does not have any direct communication links to its neighboring agents $j\in\mathcal{N}_{i}$. Moreover, the \textit{future} input $u_{j}(t_{j}^{l_{j}+1(t_{i}^{l_{i}})})$ is not available to the agent $i$ at time $t_{i}^{l_{i}}$.
In this paper, we employ the zero input response as the prediction of $x_{j}(t)$ for $t>t_{j}^{l_{j}+1}$. In Lemma~\ref{lem:1}, we will derive a bound on the uncertainty of the unknown input $u_{j}(t_{j}^{l_{j}+1})$.
\end{rem}

Define the input error by 
\begin{equation}
\label{eq:uitilde=ui-zi}
\tilde{u}_{i}(t):=u_{i}(t)-z_{i}(t).
\end{equation}
and $
\tilde{u}:=
[\tilde{u}_{1}^{\sf T},\ldots,
\tilde{u}_{N}^{\sf T}]^{\sf T}.$
Then, the collective dynamics of the agents with the above control law is expressed as
\begin{equation}
\label{eq:mascollective}
\dot{x}(t)=\mathcal{A}x+\mathcal{B}\tilde{u}(t),
\end{equation}
where $\mathcal{A}:=I_{N}\otimes A-L\otimes (BF),~\mathcal{B}:=I_{N}\otimes B$. It is easily verified that 
$
\mathcal{A}=
(U\otimes I_{n})
\mbox{diag}(
0,\check{\mathcal{A}})
(U\otimes I_{n})^{-1},
$
where 
\begin{align}
\label{eq:def-of-Acheck}
\check{\mathcal{A}}:=I_{N-1}\otimes A-\check{L}\otimes (BF).
\end{align}
As described later, the stability of $\check{\mathcal{A}}$ plays a crucial role in the design of the synchronizing control law.
 
By using a suitable nonsingular matrix $\hat{U}\in\mathds{C}^{(N-1)\times(N-1)}$, the Jordan canonical form of $\check{L}$ can be expressed as $\hat{U}^{-1}\check{L}\hat{U}$ which has a block bidiagonal structure whose diagonal elements are given by $\lambda_{2},\ldots,\lambda_{N}$, and
the off-diagonal entries are equal to $0$ or $1$ according to the geometric multiplicities of the eigenvalues. Then, we have 
\begin{align*}
&(\hat{U}\otimes I_n)^{-1}\check{\mathcal{A}}(\hat{U}\otimes I_n)
\\&= 
\begin{bmatrix} 
A- \lambda_2BF & \ast &  0  &  \cdots & 0 \\
0 & A-\lambda_3BF & \ast & \ddots & \vdots \\
0 &  0  & \ddots & \ddots &  0 \\
\vdots &  & \ddots & A-\lambda_{N-1}BF & \ast \\
0 & \cdots & \cdots & 0 & A-\lambda_N BF
\end{bmatrix},
\end{align*}
\normalsize
where $\ast$ denotes the irrelevant block entries. Since the RHS has also a block bidiagonal structure, we see that $\check{\mathcal{A}}$ is Hurwitz stable if and only if $A-\lambda_{i}BF$ is Hurwitz stable for $i=2,\ldots,N$. Based on the above observations, we make the following assumption.
\begin{assupm}
\label{assupm:Hurwitz-ind}
The matrices 
$A-\lambda_{i}BF,~i=2,\ldots,N$
are Hurwitz stable.
\end{assupm}
We employ the time-dependent threshold function $s(t)$$\in\mathds{R}$ in the following form
\begin{equation}
\label{eq:definitio-of-st}
s(t):=s_{\infty}+(s_{0}-s_{\infty})\mathrm{e}^{-\lambda_{s}t},~s_{0}\geq s_{\infty}>0, \lambda_{s}>0
\end{equation}
where $s_{0}$, $s_{\infty}$, and $\lambda_{s}$ are the constants determined by a designer. This type of the threshold function is frequently used in both standard event-triggered control methods and cloud-based self-triggered methods \citep{Almeida,Adaldo18}. Especially, the choice of parameter $\lambda_{s}$ affects the frequency of the accesses and convergence speed of the closed-loop system, $s_{0}$ affects the access frequency for some initial time period, and $s_{\infty}$ affects the bound on the steady-state synchronization errors (Also refer to Algorithm~\ref{alg1}). Although this threshold function must be designed in a centralized fashion, it is not restrictive in practical situations. Actually, if we do not care about the control performance, it is not so difficult to embed a certain common $s(t)$ to the agents like a clock.
\color{black}

\color{black}
The next lemma states that if the input error \eqref{eq:uitilde=ui-zi} is smaller than or equal to the threshold \eqref{eq:definitio-of-st}, we can bound the synchronization error and the future input by some functions respectively. More precisely, these bounds play crucial roles in the design of the triggering function in the access rule and convergence analysis.
\color{black}

\begin{lemmalemma}
\label{lem:1}
Assume that
\begin{equation}
\label{eq:ui-strel}
\|\tilde{u}_{i}(t)\|\leq s(t)
\end{equation}
is satisfied for all $t\in\left[t_{0},t_{f}\right[,~i\in\mathcal{V}$, $t_{0}<t_{f}$,  where $t_{0}$ and $t_{f}$ are arbitrary. Then, the following inequalities hold.
\begin{align}
\label{eq:delta<eta}
\displaystyle \|\delta(t)\|&\leq \eta(t)~~\forall t\in\left[t_{0},t_{f}\right[,
\\
\label{eq:|uj|<mu}
\|u_{j}(t)\|&\leq\mu_{j}(t)~~\forall j\in\mathcal{V}~~\forall t\in\left[t_{0},t_{f}\right[,
\end{align}
where the scaler-valued functions $\eta(t)$ and $\mu_{j}(t)$ are defined by
\begin{align}
\label{eq:eta-def}
\eta(t)&:= \kappa\mathrm{e}^{-\lambda(t-t_0)}\eta_{0}
+
\kappa \sqrt{N}\|\mathcal{B}'\|
\int_{t_{0}}^{t} \mathrm{e}^{-\lambda(t-\tau)}s(\tau)\mathrm{d}\tau,\\
\label{eq:muj-eta-st}
\mu_{j}(t)&:= \beta_{j}\eta(t)+s(t),
\end{align}
with $\beta_{i}:=\|[L\otimes F]_{i}\|$, $\mathcal{B}':=(I_{N}-\mathds{1}_N\mathds{\phi}^{\sf T})\otimes B$, and $\eta_{0}\geq\|\delta(t_{0})\|$ respectively. $\lambda>0$ and $\kappa>0$ are 
constants satisfying
\begin{equation}
\label{eq:pascoal-lemma}
\|\mathrm{e}^{\mathcal{A}t}v\|\leq \kappa \mathrm{e}^{-\lambda t}\|v\|,
\end{equation}
with the vector $v$ such that $(\phi^{\sf T}\otimes I_{n})v=0$ $\forall t\in\left[t_{0},t_{f}\right[$.
\end{lemmalemma}

\begin{pf}
It follows from \eqref{eq:def-alpha}, \eqref{eq:mascollective} and $\phi^{\sf T}L=0$ that 
$
\dot{\alpha}(t)=
A\alpha(t)+(\phi^{\sf T}\otimes B)\tilde{u}(t).
$
Then the dynamics of $\delta$ in \eqref{eq:def-delta} is expressed as 
\begin{align}
\dot{\delta}(t)&=\dot{x}(t)-\mathds{1}_{N}\otimes\dot{\alpha}(t)
=\mathcal{A}\delta(t)+\color{black}\mathcal{B}'\color{black}\tilde{u}(t).
\label{eq:delta-dynamics} 
\end{align}
By solving \eqref{eq:delta-dynamics}, the time evolution of $\delta(t)$ from $t_{0}$ is given by
\begin{equation}
\label{eq:delta-timeevol}
\delta(t)=\mathrm{e}^{\mathcal{A}(t-t_{0})}\delta(t_{0})+\int_{t_{0}}^{t}
\mathrm{e}^{\mathcal{A}(t-\tau)}\mathcal{B}'\tilde{u}(\tau)\mathrm{d}\tau.
\end{equation}
By a technique similar to Lemma 1 in \cite{Almeida}, it turns out that, if $\check{\mathcal{A}}$ is Hurwitz stable and $v$ satisfies $(\phi^{\sf T}\otimes I_n)v=0$, there exist $\lambda>0$ and $\kappa>0$ satisfying \eqref{eq:pascoal-lemma}. It should be noted that $\delta$ and $\mathcal{B'}\tilde{u}$ meet the above condition for $v$. We have to design the gain $F$ so that $\check{\mathcal{A}}$ is Hurwitz stable (see Remark \ref{rem:A-Hurwitz} for more detail).

It then follows that
$
\|\delta(t)\|
\leq \kappa \mathrm{e}^{-\lambda (t-t_{0})}\|\delta(t_{0})\|
+\int_{t_{0}}^{t}\kappa\mathrm{e}^{-\lambda(t-\tau)}\|\mathcal{B}'\|\|\tilde{u}(\tau)\|\mathrm{d}\tau.
$
By using \eqref{eq:ui-strel}, we further bound $\delta(t)$ as
$\|\delta(t)\|\leq\kappa\mathrm{e}^{-\lambda (t-t_{0})}\|\delta({t_{0}})\|
+
\kappa \sqrt{N}\|\mathcal{B}'\|
\int_{t_{0}}^{t} \mathrm{e}^{-\lambda(t-\tau)}s(\tau)\mathrm{d}\tau.
$
Let $\eta_{0}$ be a constant which satisfies $\eta_{0}>\|\delta(t_{0})\|$. Then, by comparing the RHS of the above inequality with \eqref{eq:eta-def}, we obtain the inequality \eqref{eq:delta<eta}.

It remains to prove the second inequality \eqref{eq:|uj|<mu}. Recall that the relative state feedback control law can be expressed as $z_{i}(t)=F\sum_{j\in\mathcal{N}_{i}}(\delta_{j}(t)-\delta_{i}(t))$. 
Thus, $z_{i}(t)$ is represented by
\begin{equation}
\label{eq:zi=cLidelta}
z_{i}(t)=-[L\otimes F]_{i}\delta(t),
\end{equation}
and taking norms of both sides leads to $
\|z_{i}(t)\|\leq\beta_{i}\|\delta(t)\|$.
Therefore, we obtain
\begin{align}
\|u_{i}(t)\|&\leq \|z_{i}(t)\|\color{black}+ \|\tilde{u}_{i}(t)\|\leq \beta_{i}\eta(t)\color{black}+s(t).
\label{eq:mu-iineq}
\end{align}Recall from \eqref{eq:muj-eta-st} that the RHS of \eqref{eq:mu-iineq} is equal to $\mu_{i}(t)$. By replacing the suffix $i$ to $j$, it concludes the proof.
\qed
\end{pf}
\begin{rem}[Choice of the feedback gains]
\label{rem:A-Hurwitz}
As described above, the feedback gain $F$ should be chosen so that the matrix $\check{\mathcal{A}}$ is Hurwitz stable. In this remark, we summarize the design procedure of $F$. Due to the structure of $\check{\mathcal{A}}$ defined in \eqref{eq:def-of-Acheck}, if all the matrices  
$
A-\lambda_{i}BF,~i=2,\ldots,N, 
$
are Hurwitz stable, then $\check{\mathcal{A}}$ is so. To make $\check{\mathcal{A}}$ Hurwitz stable, one can choose $F$ as
\begin{align}
\label{eq: choice F}
F=B^{\sf T}P,~~
\rho\geq\frac{1}{2\mathrm{Re}(\lambda_{i})},~i=2,\ldots,N,
\end{align}
where $P$$\in\mathds{R}^{n\times n}$ is the positive symmetric solution to the following algebraic Riccati inequality
\begin{align}
\label{eq:MatrixInequality}
\color{black}
A^{\sf T}P+PA-\rho^{-1}PBB^{\sf T}P\prec0.
\color{black}
\end{align} 
Actually, $A-\lambda_{i}BF$ with $F$ defined in \eqref{eq: choice F} satisfies 
$(A-\lambda_{i}BB^{\sf T}P)^{*}P+P(A-\lambda_{i}BB^{\sf T}P)
=A^{\sf T}P+PA-2\mathrm{Re}(\lambda_{i})PBB^{\sf T}P\prec0
,~i=2,\ldots,N.
$ This implies that $A-\lambda_{i} BF,~i=2,\ldots,N$ are Hurwitz stable.
Note that this approach is the same as the conventional multi-agent synchronization, such as \cite{Yang}.
\end{rem}

Next, we introduce the triggering rule based on the effects of the ZOH  and the unknown future inputs, and show that the input error is always smaller than or equal to the threshold. Define the scaler-valued function $\sigma_{i}^{l_{i}}(t)=f_{i}^{l_{i}}(t)+g_{i}^{l_{i}}(t)$ with $f_{i}^{l_{i}}(t)$ in \eqref{eq:def-of-fik} aiming at estimating the ZOH effect, and $g_{i}^{l_{i}}(t)$ in \eqref{eq:defsigma}
aiming at estimating the unknown future inputs, respectively.
\begin{align}
\label{eq:def-of-fik}
f_{i}^{l_{i}}(t)&:=\left\|F
\sum_{j\in\mathcal{N}_{i}}(\hat{x}_{j}(t)-x_{i}(t))-u_{i}(t)\right\|,
\\
g_{i}^{l_{i}}(t)&:=\|B\|\|F\|\kappa_{\theta}\sum_{j\in\mathcal{N}_{i}'(t)}\int_{t^{l_{j}+1}_{j}}^{t}\mathrm{e}^{\theta(t-\tau)}\color{black}\mu_{j}(\tau)\mathrm{d}\tau.
\label{eq:defsigma}
\end{align}
The set $\mathcal{N}_{i}'(t)$ consists of the $i$-th agent's neighbors whose control inputs are not available at time $t$, and the constants $\kappa_{\theta}\geq0$ and $\theta\in\mathds
{R}$ are chosen so that the following inequality is satisfied (Also refer to Remark \ref{rem:theta-choice}).
\begin{align}
\label{eq:theta-bound}
\|\mathrm{e}^{At}\|\leq \kappa_{\theta}\mathrm{e}^{\theta t}~~\forall t\geq0.
\end{align}
\begin{lemmalemma}
Assume that the next access time to the cloud $t_{i}^{l_{i}+1}$ is determined according to the self-triggering rule 
\begin{equation}
\label{eq:trgrule}
t_{i}^{l_{i}+1}=\inf\{t>t_{i}^{l_{i}}~|~\sigma_{i}^{l_{i}}(t)\geq s(t)\}.
\end{equation}
Suppose that the closed-loop system under the triggering rule \eqref{eq:trgrule} is well-defined on the interval $\left[0,T^{\#}\right[$ with the accesses $\{t_{i}^{l_{i}}\}_{l_{i}\in\mathds{A}_{i}},~\mathds{A}_{i}\subseteq\mathds{Z}_{\geq 0}$. Then, the inequality $
\|\tilde{u}_{i}(t)\|\leq s(t)~~\forall t \in \left[0, T^{\#}\right[,~\forall i \in \mathcal{V}
$ is satisfied. 
\label{lem:2}
\end{lemmalemma}
\begin{pf}
We prove this lemma by taking a similar approach to \cite{Adaldo17,Adaldo18}. The prediction $\hat{x}_{j}(t)$ by the agent $i$ at time $t_{i}^{l_{i}}$ is divided into the following two cases. Define $t_{j}^{l_{j}}$ as the latest access time of the agent $j$ before $t_{i}^{l_{i}}$. For $t\leq t_{j}^{l_{j}+1}$, the control input $u_{j}(t)=z_{j}(t_{j}^{l_{j}})$ is available, and hence $x_{j}(t) (=\hat{x}_{j}(t))$ can be exactly computed. On the other hand, for $t>t_{j}^{l_{j}+1}$, the input $u_{j}(t_{j}^{l_{j}+1})$ is not available. Thus, we can decompose $x_{j}$ into 
\begin{equation}
\label{eq:xjfort>tjh+1}
x_{j}(t)=\hat{x}_{j}(t)+\int_{t_{j}^{l_{j}+1}}^{t}\mathrm{e}^{A(t-\tau)}Bu_{j}(\tau)\mathrm{d}\tau.
\end{equation}
Note that the second term on the RHS is not available to the agent $i$.

We can express the input error $\|\tilde{u}_{i}(t)\|$ by
\begin{equation*}
\|\tilde{u}_{i}(t)\|=
\Biggl\|u_{i}(t)-F\sum_{j\in\mathcal{N}_{i}}(\hat{x}_{j}(t)-x_{i}(t))-F\sum_{j\in\mathcal{N}_{i}'(t)}\int_{t_{j}^{l_{j}+1}}^{t}\mathrm{e}^{A(t-\tau)}Bu_{j}(\tau)\mathrm{d}\tau\Biggr\|,
\end{equation*}
where we used \eqref{eq:ut}, \eqref{eq:predictionsofxj},
\eqref{eq:uitilde=ui-zi}, and \eqref{eq:xjfort>tjh+1}. Application of the triangle inequality to the above equation yields
\begin{align}
\|\tilde{u}_{i}(t)\|
\leq \left\|u_{i}-F\sum_{j\in\mathcal{N}_{i}}(\hat{x}_{j}(t)-x_{i}(t))\right\|
+
\|B\|\|F\|\kappa_{\theta}\sum_{j\in\mathcal{N}_{i}'(t)}\int_{t_{j}^{l_{j}+1}}^{t}\mathrm{e}^{\theta(t-\tau)}\|u_{j}(\tau)\|\mathrm{d}\tau.
\notag
\end{align}
Let us consider the situation that the agent $i$ satisfies $\|\tilde{u}_{i}(t)\|>\sigma_{i}^{l_{i}}(t)$ at $t\in ]t_{i}^{l_{i}},t_{i}^{l_{i}+1}[~$. Then there exists $j\in\mathcal{V}, \tau\in ]t_{j}^{l_{j}+1},t[$ such that $\|u_{j}(\tau)\|\geq \mu_{j}(\tau)$. However, by contraposition of Lemma 1, this condition means there exists $ \tau'\in ]t_{j}^{l_{j}+1},t [$,~$j'\in\mathcal{V}$ satisfying $\|\tilde{u}_{j'}(\tau')\|>s(\tau')$. This implies $\|\tilde{u}_{j'}(\tau')\|>\sigma_{j'}^{l_{j'}(\tau')}(\tau')$. Therefore, $\|\tilde{u}_{i}(t)\|\leq\sigma_{i}^{l_{i}}(t)$ will not be violated unless another agent violates the condition. We can take $\|\tilde{u}_{i}(0)\|=0$, thus we can conclude the statement of Lemma \ref{lem:2}.
\qed
\end{pf}
It may be remarked that the closed-loop system under the triggering rule \eqref{eq:trgrule} is actually given by 
\begin{equation}
\label{eq:delta-timeevol-2}
\delta(t)=\mathrm{e}^{\mathcal{A}t}\delta(0)+\int_{0}^{t}
\mathrm{e}^{\mathcal{A}(t-\tau)}\mathcal{B}'\tilde{u}(\tau)\mathrm{d}\tau,~t\in\left[0,T^{\#}\right[,
\end{equation}
with \eqref{eq:uitilde=ui-zi}, \eqref{eq:trgrule}
and \eqref{eq:delta-timeevol}  according to Lemma \ref{lem:1}.
\begin{rem}
\label{rem:theta-choice}
It is seen from the construction of \eqref{eq:trgrule} that the parameter $\theta$ affects the access frequency. (As discussed in the next sub-section, it also affects the lower bound of the access interval).  To be more precise, larger $\theta$ and $\kappa_{\theta}>0$ lead to more frequent accesses and smaller lower bound of the access interval. Therefore, $\theta$ should be tightly calculated for accurate estimation of the access frequency and the lower bound of the access interval. To obtain a tight bound on the matrix exponential in \eqref{eq:theta-bound}, one can solve the following convex programmings $\theta=\inf_{\mathcal{P}\succ0,\vartheta>0}\{\vartheta~|~A^{\sf T}\mathcal{P}+\mathcal{P}A\prec2\vartheta \mathcal{P}\}$ with $\mathcal{P}\in\mathds{R}^{n\times n}$, and 
$
\rho({\theta})=\inf_{\rho, \mathcal{P}}\{\rho |~I_{n}\prec\color{black}\mathcal{P}\prec\color{black}\rho I_{n},A^{\sf T}\mathcal{P}+\mathcal{P}A\prec2\theta\color{black}\mathcal{P}\}.
$
We then obtain the bound $\|\mathrm{e}^{At}\|\leq \sqrt{\rho(\theta)}\mathrm{e}^{\theta t}$ $\forall t\geq 0$. Especially, if the matrix $A$ is diagonalizable, the above infimum is achieved by $\theta= \max_{i}\mathrm{Re}(\mathrm{eig}(A))$ and $A^{\sf T}\mathcal{P}+\mathcal{P}A\preceq 2\theta \mathcal{P}$ in the above condition. We can also obtain tight parameters $\kappa$ and $\lambda$ in \eqref{eq:pascoal-lemma} by applying the same technique to $\check{\mathcal{A}}$.
\end{rem}
\begin{rem}
\label{rem:5}
Note that there is a possibility that we can take the next access time $t_{i}^{l_{i}+1}$ arbitrarily large in the triggering rule \eqref{eq:trgrule}. This means that the desired error bound can be achieved without any information exchanges with its neighbors after a certain access time. In this case, the next access time can be set to $t_{i}^{l_{i}+1}=\infty$ which means that the agent will not access the cloud any more after $t_{i}^{l_{i}}$, although arbitrary finite $t_{i}^{l_{i}+1}$ is admissible. 
\end{rem}
\begin{rem}
It may be noted that our proposed access rule actually can be implemented in a self-triggered fashion as can be seen from \eqref{eq:trgrule}. Namely, each agent can calculate its next access time $t_{i}^{l_{i}+1}$ at every access time $t_{i}^{l_{i}}$ by using the predictions of $x_{i}(t)$ and $x_{j}(t)$ based on information in the cloud at $t_{i}^{l_{i}}$.
\end{rem}
\subsection{Guarantee of Non-Zeno Behavior}
In this sub-section, we will prove that the closed-loop system does not exhibit any Zeno behavior, namely, the sequence of the triggering times $\{t_{i}^{l_{i}}\}_{l_{i}\in\mathds{A}_{i}}$ does not have any accumulation points.
\begin{lemmalemma}
\label{lem:Zeno}
The closed-loop system \eqref{eq:mascollective} under the triggering rule \eqref{eq:trgrule} does not exhibit any Zeno behavior.
\end{lemmalemma}
\begin{pf}
To prove this Lemma, we will show that the inter-event period $t_{i}^{l_{i}+1}-t_{i}^{l_{i}}$ has a positive lower bound $\tau_{\color{black}i\color{black}}^{\star}$. Moreover, we will argue that  the closed-loop system is well-defined on $[0,\infty[$, i.e., $T^{\#}=\infty$. As mentioned in \citet{Liu,Liu2}, the following three cases are considerable: a) $\lim_{l_{i}\rightarrow\infty}t_{i}^{l_{i}}<\infty$, b) $\lim_{l_{i}\rightarrow\infty} t_{i}^{l_{i}}=\infty$ and $T^{\#}=\infty$, and c) $\mathds{A}_{i}$ is a finite set. Indeed, the case a) is the undesired Zeno behavior and  should be excluded.

Suppose that $t\in[t_{i}^{l_{i}},t_{i}^{l_{i}+1}[,~l_{i}\in\mathds{A}_{i}$. Firstly, let us calculate the upper bound of $f_{i}^{l_{i}}(t)$. Recall that $f_{i}^{l_{i}}(t)$ in \eqref{eq:def-of-fik} can be expressed as
\begin{multline}
\label{eq:fikwithuhat}
f_{i}^{l_{i}}(t)=\Biggl\|
F\sum_{j\in\mathcal{N}_{i}}(\hat{x}_{j}(t_{i}^{l_{i}})-x_{i}(t_{i}^{l_{i}})
-F\sum_{j\in\mathcal{N}_{i}}
\biggr\{
\mathrm{e}^{A(t-t_{i}^{l_{i}})}\hat{x}_{j}(t_{i}^{l_{i}})+\int_{t_{i}^{l_{i}}}^{t}\mathrm{e}^{A(t-\tau)}B\hat{u}_{j}(\tau)\mathrm{d}\tau\\
-\bigl(\mathrm{e}^{A(t-t_{i}^{l_{i}})}x_{i}(t_{i}^{l_{i}})+\int_{t_{i}^{l_{i}}}^{t}
\mathrm{e}^{A(t-\tau)}Bu_{i}(\tau)\mathrm{d}\tau\bigr)
\biggl\}\Biggr\|,
\end{multline}
with  $
\hat{u}_{j}(t):=\begin{cases} u_{j}(t) & t\in[t_{i}^{l_{i}},t_{j}^{l_{j}+1}[\\
0 & t\geq t_{j}^{l_{j}+1},
\end{cases}
$
, where $t_{j}^{l_{j}+1}$ denotes the next triggering time of the agent $j$. Hereafter, we prove this lemma for each sign of $\theta$ in \eqref{eq:theta-bound}.

By applying the triangle inequality to \eqref{eq:fikwithuhat}, 
\begin{align}
f_{i}^{l_{i}}(t)&\leq
\|\mathrm{e}^{A(t-t_{i}^{l_{i}})}-I\|\Biggl\|F\sum_{j\in\mathcal{N}_{i}}(\hat{x}_{j}(t_{i}^{l_{i}})-x_{i}(t_{i}^{l_{i}}))\Biggr\|\notag
\\
&+
\|F\|\sum_{j\in\mathcal{N}_{i}}\Biggl(
\int_{t_{i}^{l_{i}}}^{t}\|\mathrm{e}^{A(t-\tau)}\|\|B\|\|\hat{u}_{j}(\tau)\|\mathrm{d}\tau+
\int_{t_{i}^{l_{i}}}^{t}\|\mathrm{e}^{A(t-\tau)}\|\|B\|\|u_{i}(\tau)\|\mathrm{d}\tau
\Biggr).
\label{eq:fik-upbou1}
\end{align}
Recall that
\begin{align*}
\bigl\|\mathrm{e}^{A(t-t_{i}^{l_{i}})}-I\bigr\|
&=\Biggl\|\int_{0}^{t-t_{i}^{l_{i}}}A\mathrm{e}^{A\tau}\mathrm{d}\tau\Biggr\|\leq \int_{0}^{t-t_{i}^{l_{i}}}\|A\|\|\mathrm{e}^{A\tau}\|\mathrm{d}\tau\\
&=
\begin{cases}
\kappa_{\theta}\|A\|(t-t_{i}^{l_{i}}) & \theta=0\\
\displaystyle\kappa_{\theta}\|A\|(\mathrm{e}^{\theta(t-t_{i}^{l_{i}})}-1)/\theta & \theta\neq0
\end{cases}.
\end{align*}
Firstly, consider the case of $\theta\neq0$. The first term on the RHS of \eqref{eq:fik-upbou1} is further bounded by
\begin{align}
\label{eq:fik1stupbou}
&\mbox{[1st term on RHS of \eqref{eq:fik-upbou1}]}\notag\\
&\leq\frac{\kappa_{\theta}\|A\|}{\theta}(\mathrm{e}^{\theta(t-t_{i}^{l_{i}})}-1)\|[L\otimes F]_{i}\delta(t_{i}^{l_{i}})\|\notag\\
&\color{black}\leq \frac{\kappa_{\theta}\beta_{i}\overline{\eta}\|A\|}{\theta}(\mathrm{e}^{\theta(t-t_{i}^{l_{i}})}-1),
\end{align}
where $\overline{\eta}$ is an upper bound on $\eta(t)$ in \eqref{eq:eta-def}, which is the bound on the closed-loop dynamics $\delta(t)$ in \eqref{eq:delta-timeevol}. The first inequality is obtained by using \eqref{eq:ut}, \eqref{eq:predictionsofxj} and \eqref{eq:zi=cLidelta}.  
Similarly, the second term of \eqref{eq:fik-upbou1} is also bounded by 
\begin{align}
&\mbox{[2nd term on RHS of \eqref{eq:fik-upbou1}]}\notag\\
&\leq \|B\|\|F\|\kappa_{\theta}\sum_{j\in\mathcal{N}_{i}}\Biggl(\int_{t_{i}^{l_{i}}}^{t}\mathrm{e}^{\theta(t-\tau)}\mu_{j}(\tau)\mathrm{d}\tau+
\int_{t_{i}^{l_{i}}}^{t}\mathrm{e}^{\theta(t-\tau)}\mu_{i}(\tau)\mathrm{d}\tau\Biggr)\notag\\
&\leq \|B\|\|F\|\kappa_{\theta}\sum_{j\in\mathcal{N}_{i}}
\Biggl(\int_{t_{i}^{l_{i}}}^{t}\mathrm{e}^{\theta(t-\tau)}(\beta_{j}\overline{\eta}+s_{0})\mathrm{d}\tau
+
\int_{t_{i}^{l_{i}}}^{t}\mathrm{e}^{\theta(t-\tau)}(\beta_{i}\overline{\eta}+s_{0})\mathrm{d}\tau
\Biggr)\notag
\\
&= \|B\|\|F\|\kappa_\theta\sum_{j\in\mathcal{N}_{i}}
\Biggl\{
\frac{\beta_{j}\overline{\eta}+s_{0}}{\theta}(\mathrm{e}^{\theta(t-t_{i}^{l_{i}})}-1)+
\frac{\beta_{i}\overline{\eta}+s_{0}}{\theta}(\mathrm{e}^{\theta(t-t_{i}^{l_{i}})}
-1)\Biggr\}.
\label{eq:fik2ndupbou}
\end{align}
\normalsize
Note that the first inequality is obtained by \eqref{eq:|uj|<mu} in Lemma~\ref{lem:1}, and the second inequality is obtained by \eqref{eq:muj-eta-st}. Also, $g_{i}^{l_{i}}(t)$ is bounded by  
\begin{align}
g_{i}^{l_{i}}(t)&\leq \|B\|\|F\|\kappa_{\theta}\sum_{j\in\mathcal{N}_{i}}\int_{t_{i}^{l_{i}}}^{t}\mathrm{e}^{\theta(t-\tau)}\mu_{j}(\tau)\mathrm{d}\tau
\notag
\\
&\leq \|B\|\|F\|\kappa_{\theta}\sum_{j\in\mathcal{N}_{i}}(\beta_{j}\overline{\eta}+s_{0})\int_{t_{i}^{l_{i}}}^{t}\mathrm{e}^{\theta(t-\tau)}\mathrm{d}\tau
\notag
\\
&=\|B\|\|F\|\kappa_{\theta}\sum_{j\in\mathcal{N}_{i}}\frac{\beta_{j}\overline{\eta}+s_{0}}{\theta}\bigl(\mathrm{e}^{\theta(t-t_{i}^{l_{i}})}-1\bigr).\color{black}
\label{eq:gbound}
\end{align}
By summing up \eqref{eq:fik1stupbou}, \eqref{eq:fik2ndupbou} and \eqref{eq:gbound}, we get
\begin{align}
\sigma_{i}^{l_{i}}(t)&=f_{i}^{l_{i}}(t)+g_{i}^{l_{i}}(t)\notag\\
&\leq\|B\|\|F\|\kappa_{\theta}
\sum_{j\in\mathcal{N}_{i}}
\frac{(\beta_{i}+2\beta_{j})\overline{\eta}+3s_{0}}{\theta}
(\mathrm{e}^{\theta(t-t_{i}^{l_{i}})}-1)\notag\\
&\mbox{~~~~~~}+\frac{\kappa_{\theta}\beta_{i}\overline{\eta}\|A\|}{\theta}(\mathrm{e}^{\theta(t-t_{i}^{l_{i}})}-1)\notag\\
&=\gamma_{i}\frac{\mathrm{e}^{\theta(t-t_{i}^{l_{i}})}-1}{\theta},
\label{eq:sigma-upperbound}
\end{align}
where $\gamma_{i}$ is the positive constant defined by
$\gamma_{i}=\|B\|\|F\|\kappa_{\theta}
\sum_{j\in\mathcal{N}_{i}}
(\beta_{i}+2\beta_{j})\overline{\eta}+3s_{0}
+\kappa_{\theta}\beta_{i}\overline{\eta}\|A\|.
$ By recalling that from \eqref{eq:definitio-of-st} $s(t)$ satisfies
$s(t)\geq s_{\infty}~~\forall t\geq t_{i}^{l_{i}},$
and combining with \eqref{eq:sigma-upperbound} the triggering condition $\sigma_{i}^{l_{i}}(t)\geq s(t)$ in \eqref{eq:trgrule} is not satisfied unless $
\gamma_{i}(\mathrm{e}^{\theta(t-t_{i}^{l_{i}})}-1)/\theta
<  s_\infty~\forall t\geq t_{i}^{l_{i}},
$ is violated. 
Let $\tau_{i}^{\star}=t-t_{i}^{l_{i}}$ be the smallest nonnegative root for 
$
\gamma_{i}(\mathrm{e}^{\theta\tau_{i}^{\star}}-1)/\theta
=s_{\infty}$.
It should be noted that the triggering condition \eqref{eq:trgrule} is not satisfied before $t_{i}^{l_{i}}+\tau_{i}^{\star}$. Hence, $\tau_{i}^{\star}$ is a lower bound on $t_{i}^{l_{i}+1}-t_{i}^{l_{i}}$ for the triggering rule \eqref{eq:trgrule}, namely $t_{i}^{l_{i}+1}-t_{i}^{l_{i}}\geq \tau_{i}^{\star}~\forall i\in\mathcal{V},~\forall l_{i}\in\mathds{A}_{i}$. To exclude any Zeno behavior, we wish to prove $\tau_{i}^{\star}>0$.

(i) In the case of $\theta>0$, $\tau_{i}^{\star}$ is explicitly given by 
\begin{align}
\label{tauistar}
\tau_{i}^{\star}=\frac{1}{\theta}\ln\left(1+\frac{\theta s_{\infty}}{\gamma_{i}}\right)>0~\forall i,~\forall l_{i}.
\end{align} 
Fig.~\ref{fig:tau-exis} illustrates the relation among $t_{i}^{l_{i}}$, $t_{i}^{l_{i}+1}$, and $\tau_{i}^{\star}$. 
\begin{figure}[htb]
\begin{center}
\includegraphics[width=7cm]{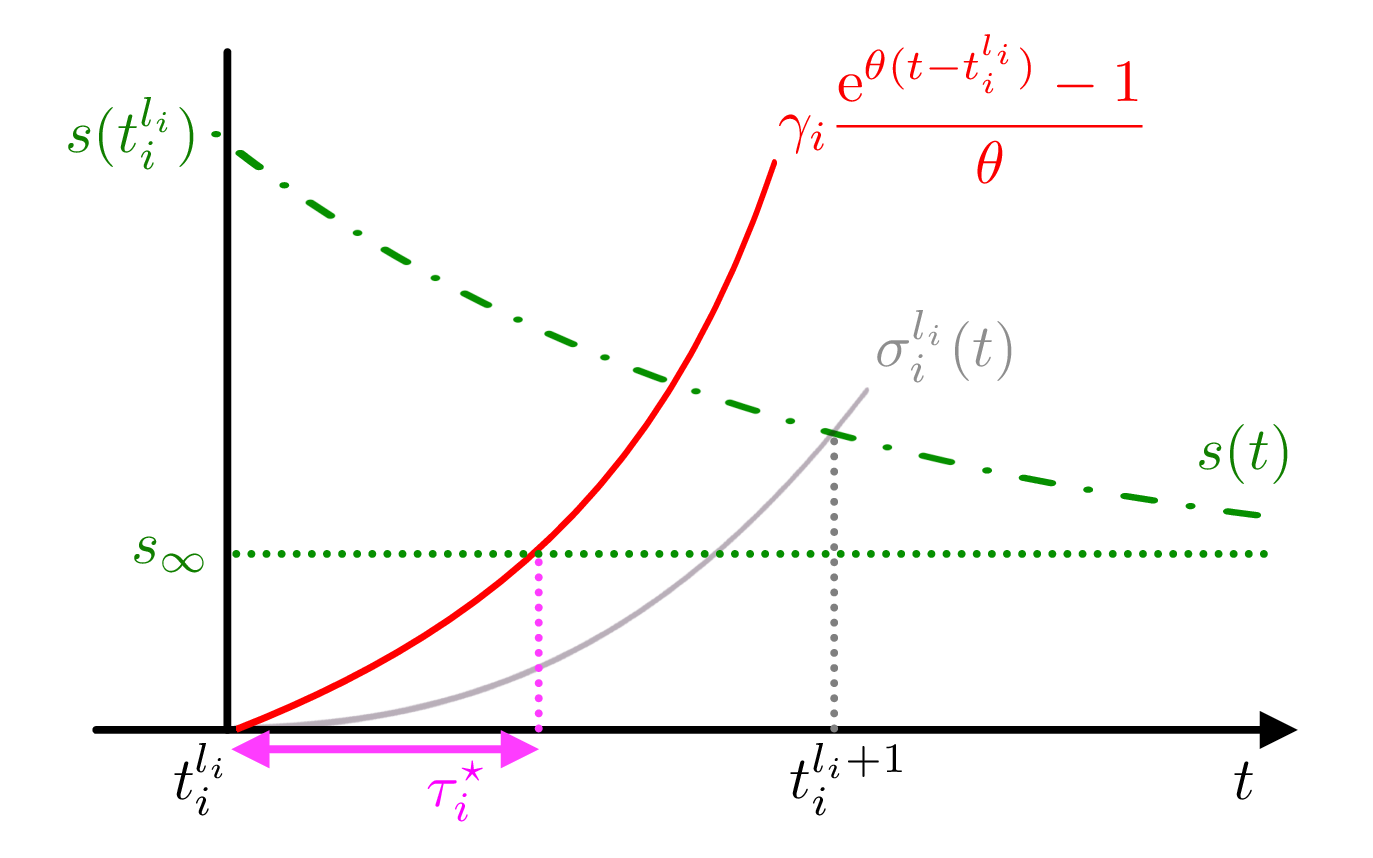}
\end{center}
\vspace{-1.3em}
\caption{Illustration of the relation among $t_{i}^{l_{i}}$, $t_{i}^{l_{i}+1}$, and $\tau_{i}^{\star}$}
\label{fig:tau-exis}
\end{figure}\\
Assume that  $\mathds{A}_{i}=\mathds{Z}_{\geq 0}$ and $\lim_{l_{i}\rightarrow \infty}t_{i}^{l_{i}}<\infty~\forall i\in\mathcal{V}$. Then, there exists a finite time $T$ satisfying $\gamma_{i}(\mathrm{e}^{\theta(T-T)}-1)/\theta=s_{\infty}$. However, such $T$ cannot exist and it contradicts \eqref{tauistar}. We thus conclude that $t_{i}^{l_{i}}\rightarrow \infty$ as $l_{i}\rightarrow\infty$, and $T^{\#}=\infty$. As another case, if $\mathds{A}_{i}=\{1,2,\ldots,\overline{l_{i}}\}~\forall i\in \mathcal{V}$ are finite sets (where $\overline{l}_{i}$ denotes a finite number of the accesses as mentioned in Remark~\ref{rem:5}), the closed-loop system is then reduced to an LTI system with continuous inputs after a certain time. This implies that the closed-loop system is well-defined on $t\in[0,\infty[$ which also means $T^{\#}=\infty$. Consequently, we can prove that $T^{\#}$ must be infinity for all the aforementioned cases.

(ii) In the case of $\theta<0$, if $\theta s_{\infty}>-\gamma_{i}$, $\tau_{i}^{\star}$ is given by \eqref{tauistar}. On the other hand, if $\theta s_{\infty}<-\gamma_{i}$, $\tau_{i}^{\star}$ does not exist. This means that $\tau_{i}^{\star}$ satisfying the triggering function does not exist. Intuitively speaking, since the dynamics of the agent is stable in the case of $\theta<0$ and $s_{\infty}$ is large, the desired tolerance can be achieved without the communication through the cloud. It should be noted that any Zeno behavior is also excluded due to the similar argument to (i).

Lastly, consider the case of $\theta=0$. We can bound $\sigma_{i}^{l_{i}}$ as $\sigma_{i}^{l_{i}}=f_{i}^{l_{i}}+g_{i}^{l_{i}}\leq\gamma_{i}(t-t_{i}^{l_{i}})$. 

Similar to the case of $\theta\neq 0$, we can get $\tau_{i}^{\star}=s_{\infty}/\gamma_{i}>0$.
It can also be proved that $T^{\#}=\infty$. We can conclude that the closed-loop system does not exhibit any Zeno behavior.
\qed
\end{pf}
\subsection{Analysis on Bounded Consensus}
In this subsection, we will study the convergence of the proposed control method to present the main claim of this paper.
\begin{thmthm}
\label{Thm1}
Let a positive constant $\eta_{0}>0$ be given. For any initial conditions $x_{i}(0)$, $i=1,\ldots,N$ which satisfies $\|\delta(0)\|<\eta_{0}$, the closed-loop system \eqref{eq:mascollective} under the triggering rule \eqref{eq:trgrule} achieves the bounded synchronization without any Zeno behavior, where the tolerance $\epsilon$ in \eqref{eq:def-boundedsynchronization} is given by $\epsilon=(\kappa\sqrt{N}\|\mathcal{B}'\|s_{\infty})/\lambda$. Moreover, the closed-loop system $\delta(t)$ is UUB with $b^{\natural}\geq\epsilon$ and $c^{\natural}\leq \eta_{0}$ in \eqref{eq:def-uub}.   
\end{thmthm}
\begin{pf}
\color{black}According to Lemmas~\ref{lem:2} and \ref{lem:Zeno}, the antecedent of Lemma~\ref{lem:1} is satisfied for all $t\geq0$. Hence, the closed-loop system $\delta(t)$ is actually given in \eqref{eq:delta-timeevol-2} on $\left[0,\infty\right[$, and thus upper bounded by $\eta(t)$ as in \eqref{eq:delta<eta}. Moreover, $\eta(t)$ in \eqref{eq:eta-def} can be written as 
\begin{align}
\label{eq:eta-timeevol}
\eta(t)=\kappa\mathrm{e}^{-\lambda t}\eta_{0}+\kappa\sqrt{N}\|\mathcal{B}'\|
\Bigl(
\frac{s_{\infty}}{\lambda}(1-\mathrm{e}^{-\lambda t})+
\frac{s_{0}-s_{\infty}}{\lambda-\lambda_{s}}(\mathrm{e}^{-\lambda_{s}t}-\mathrm{e}^{-\lambda t})
\Bigr),
\end{align}
where $\eta_{0}>\|\delta(0)\|$ by combining \eqref{eq:definitio-of-st} and \eqref{eq:eta-def}. Recall that Lemma \ref{lem:Zeno} guarantees that $\delta(t)$ is well-defined for all $t\geq 0$ without any Zeno behavior, and hence we get $\|\delta(t)\|\leq\eta(t)\longrightarrow \kappa \sqrt{N}\|\mathcal{B}'\|s_{\infty}/\lambda$ by taking the limit $t\rightarrow\infty$ in \eqref{eq:eta-timeevol}. We can immediately prove that the synchronization error is UUB. It concludes the proof.
\qed
\end{pf}
Based on the above observations, choosing smaller $s_{\infty}$ means that the bounded synchronization can be achieved with the smaller tolerance, whereas the smaller lower bound of inter-access interval is required. 
\subsection{Algorithm}
The proposed procedures for controller design and execution are shown in \textrm{Algorithms}~\ref{alg1} and  \ref{alg2}.
\begin{algorithm}[htb]
\caption{Parameter design}
\label{alg1}
\begin{enumerate}
\small
\renewcommand{\labelenumi}{\texttt{\arabic{enumi}:~}}
\item{Choose a sufficiently large $\eta_{0}$.}
\item{Compute $\|\mathcal{B}'\|,\beta_{i},~i=1,\ldots,N$.}
\item{Choose the gain $F$ satisfying Assumption~\ref{assupm:Hurwitz-ind} according to \eqref{eq: choice F} and \eqref{eq:MatrixInequality}.}
\item{Choose $\kappa_{\theta},\theta$ satisfying \eqref{eq:theta-bound}. }
\item{Choose $\kappa, \lambda$ satisfying \eqref{eq:pascoal-lemma} with $t_{0}=0$ and $t_{f}=\infty$.}
\item{Choose $s_{0}, s_{\infty}, \lambda_{s}$ in \eqref{eq:definitio-of-st}. Especially, $s_{\infty}$ is chosen so that the bounded synchronization \eqref{eq:def-boundedsynchronization} is achieved for the given $\epsilon$, namely, $s_{\infty}\leq \lambda\epsilon/\kappa\sqrt{N}\|\mathcal{B}'\|.$} 
\end{enumerate}
\end{algorithm}
\begin{algorithm}
\caption{Self-triggered synchronizing algorithm through a cloud}
\label{alg2}
\mbox{}\\
\textbf{Step 1} (Initialization)
\begin{enumerate}
\small
\renewcommand{\labelenumi}{\texttt{\arabic{enumi}:~}}
\item{Initialize $x_{i}(0),~i=1,\ldots,N$.}
\end{enumerate}
\textbf{Step 2} \\
At every time instant $t>0$, the agent $i$ executes the following steps.
\begin{enumerate}
\small
\renewcommand{\labelenumi}{\texttt{\arabic{enumi}:~}}
\item{Access the cloud at predetermined time $t_{i}^{l_{i}}$.}
\item{Compute the predictions $\hat{x}_{j},~j\in\mathcal{N}_{i}$ according to  \eqref{eq:predictionsofxj} by using the information stored in the cloud.}
\item{Compute the control input and the next access time by using \eqref{eq:ut}, \eqref{eq:trgrule}.}
\item{Save $t_{i}^{l_{i}},x_{i}(t_{i}^{l_{i}}),u_{i}(t_{i}^{l_{i}}),t_{i}^{l_{i}+1}$ to the cloud.}
\item{Disconnect from the cloud.}
\end{enumerate}
\color{black}
\end{algorithm}
\normalsize
\begin{rem}
As described in Algorithms \ref{alg1} and \ref{alg2}, the proposed algorithm requires some computation of global information in the initial setting. But, this will not be a severe drawback of this paper because most of the previous works (e.g., \cite{Adaldo17}) also need similar computations. Indeed, it suffices to choose $\eta_{0}$ sufficiently large to guarantee $\eta_{0}>\|\delta(0)\|$ without the knowledge of $\delta(0)$. In this case, the self-triggered control of Algorithm~\ref{alg2} can be fully distributed.
\end{rem}
\section{Numerical Simulation}
\begin{figure}[htb]
\begin{center}
\includegraphics[width=3.4cm]{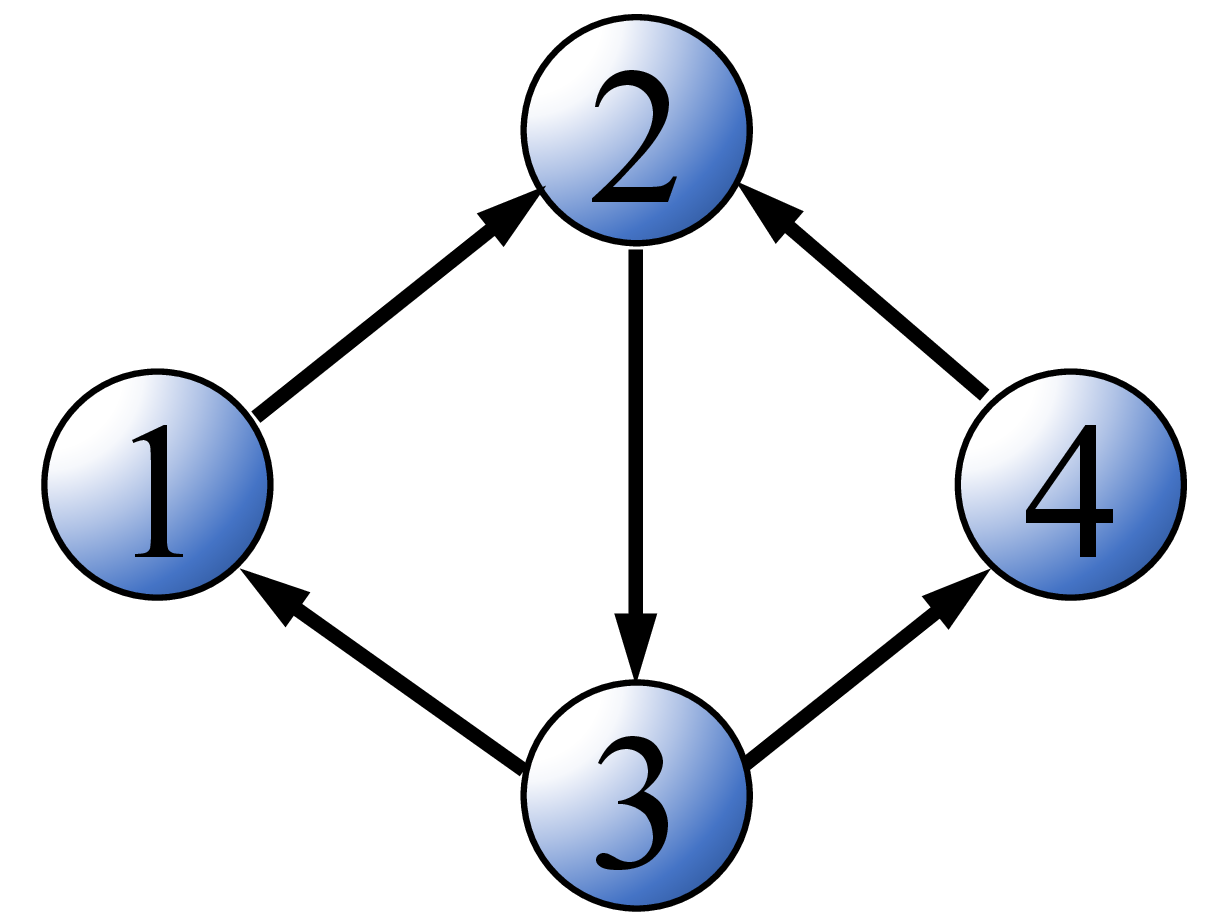}
\end{center}
\caption{Accessibility graph}
\label{fig:graph}
\end{figure}

Consider the multi-agent system consisting of $4$ agents whose dynamics is the second-order oscillator 
$\dot{x}_{i}
=
\begin{bmatrix}
0 & -0.4\\
0.4 & 0
\end{bmatrix}
x_{i}+
\begin{bmatrix}
1 & 0\\
0 & 1
\end{bmatrix}u_{i},~i=1,\ldots,4.$ The accessibility graph is depicted in Fig.~\ref{fig:graph}.
The eigenvalues of $L$ are $\mathrm{eig}(L)=\{0,1,2\pm\mathbf{j}\}$. 
\normalsize
The left eigenvector of $L$ is $\phi\color{black}=[0.2,0.2,0.4,0.2]^{\sf T}$. 
In order to find a solution of \eqref{eq:MatrixInequality}, we solve the algebraic Riccati equation 
$A^{\sf T}P+PA-\varrho^{-1}PBB^{\sf T}P+I_n=0$
with $\varrho=0.6$, and obtain
$P=\mathrm{diag}\{0.7746, 0.7746\}$.
Correspondingly, the feedback gain is obtained by 
$F=B^{\sf T}P=
\mbox{diag}\{0.7746,0.7746\}
$.
We set $\kappa_{\theta}=1$, $\theta=0$, $\lambda=0.7736, \kappa=2.3268$, and $s(t)=0.01+(1-0.01)\exp(-0.3t)$. We choose $\eta_0=15.12$. We get the lower bounds of the access intervals $\tau_{1}^{\star}=1.5102\times 10^{-4},~\tau_{2}^{\star}=9.3436\times 10^{-5},~\tau_{3}^{\star}=1.6214\times 10^{-4},~\tau_{4}^{\star}=1.2429\times 10^{-4}$[s], respectively. 

\begin{figure}[htb]
\begin{center}
\includegraphics[width=6cm]{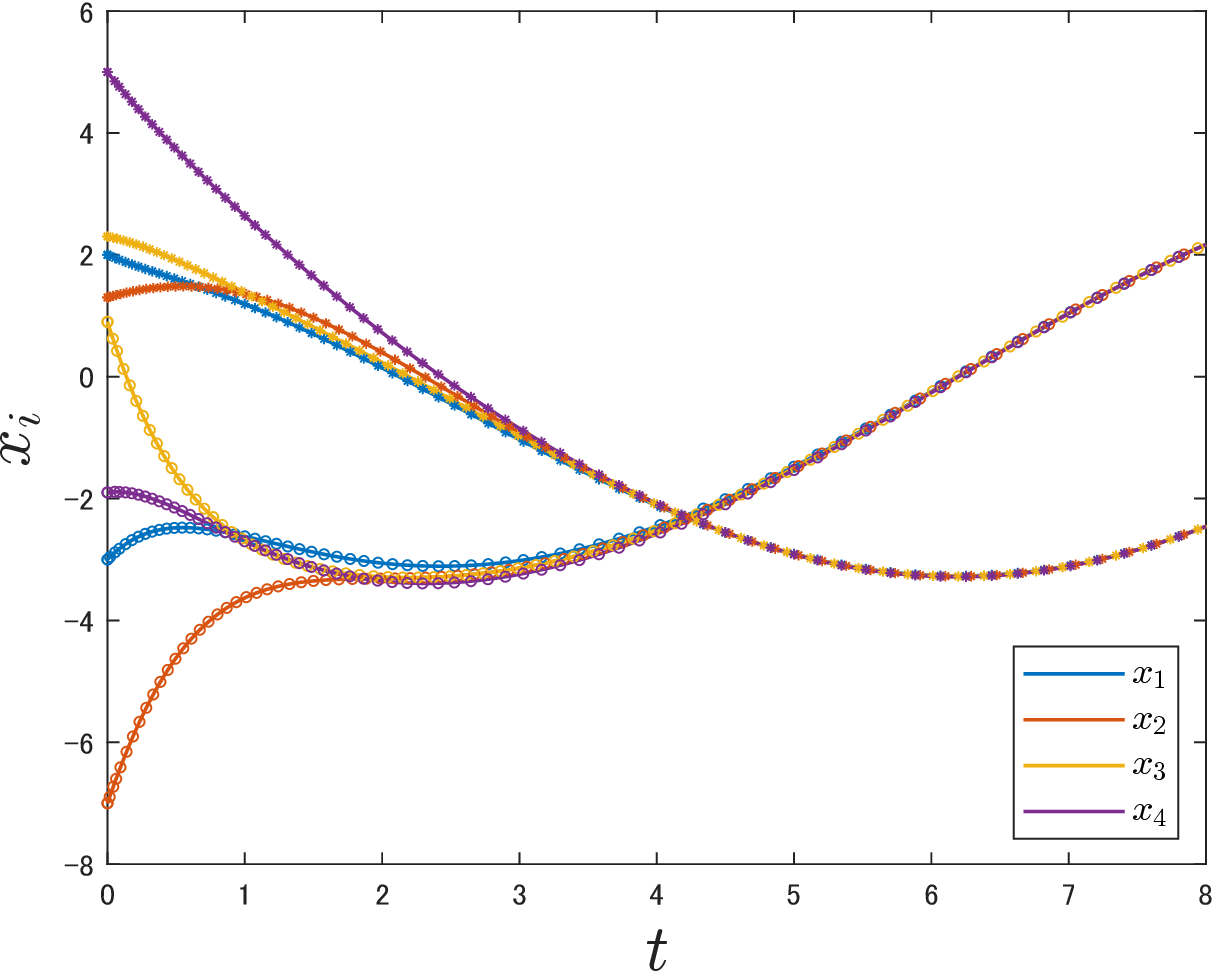}
\end{center}
\caption{States $x_{i}$}
\label{fig:trajectory}
\end{figure}

\begin{figure}[htb]
\begin{center}
\includegraphics[width=6cm]{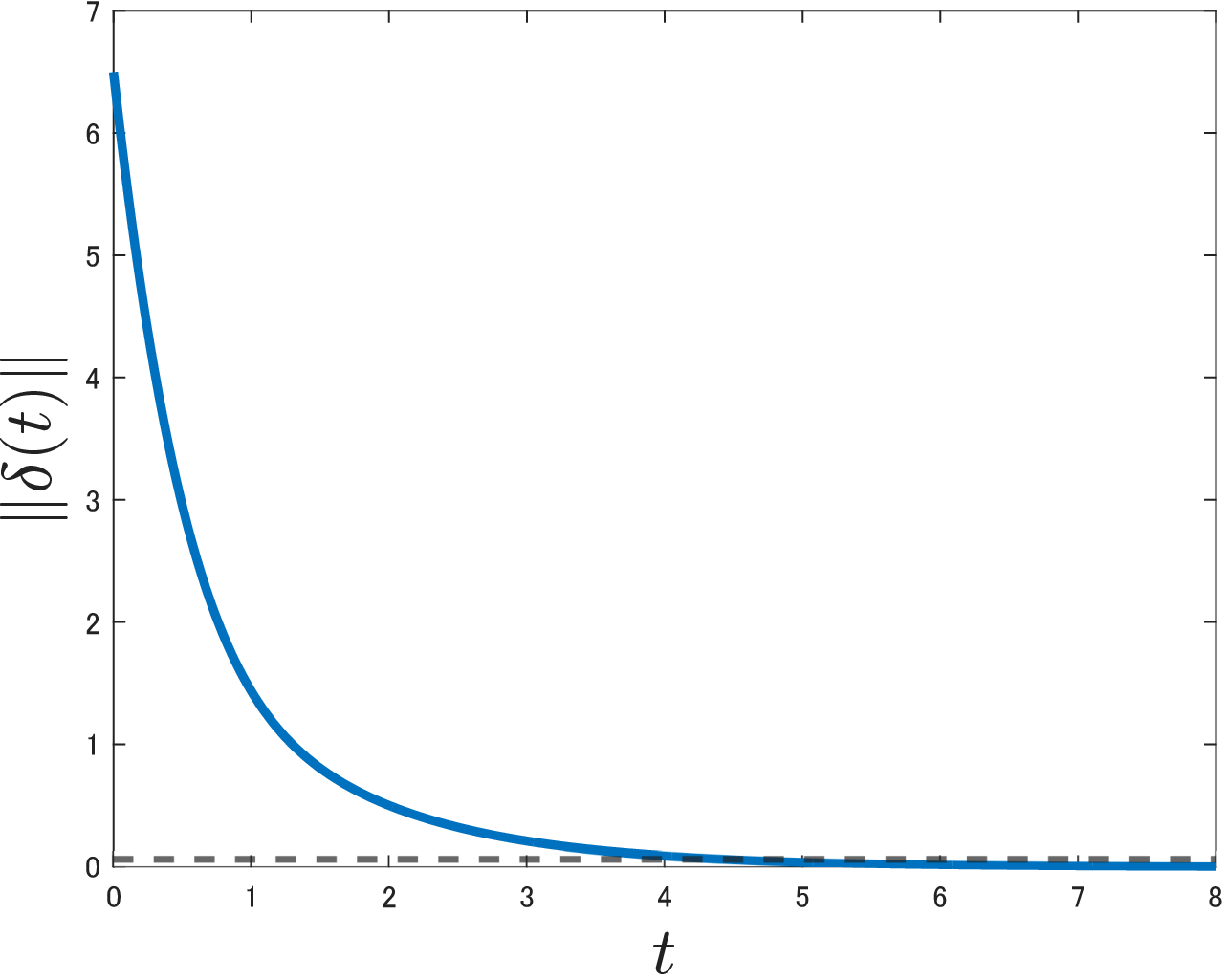}
\end{center}
\caption{Synchronization error $\|\delta(t)\|$}
\label{fig:delta}
\end{figure}

\begin{figure}[htb]
\begin{center}
\includegraphics[width=6cm]{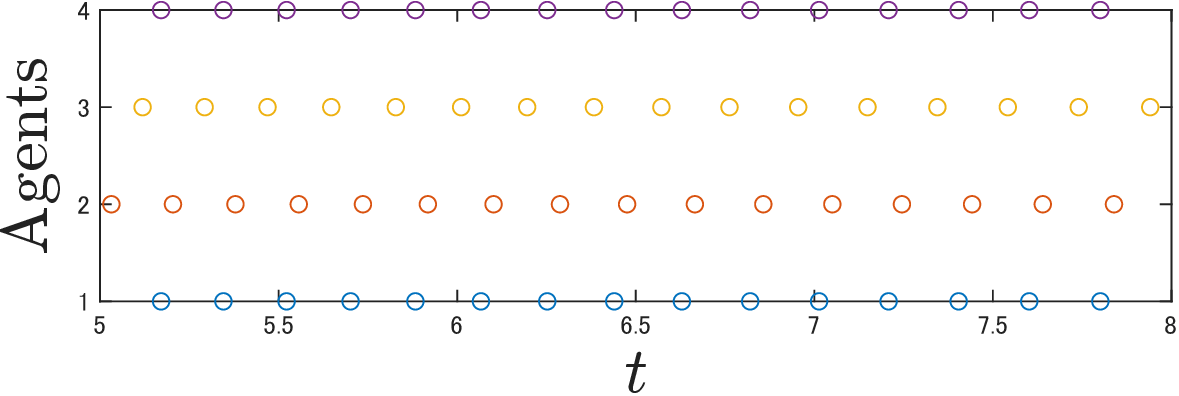}
\end{center}
\caption{Accesses within the interval from $t=5$[s] to $t=8$[s]}
\label{fig:acceses}
\end{figure}

The simulation results are shown in Figs. \ref{fig:trajectory}-\ref{fig:acceses} and Table \ref{table:cloudaccess}. We run the simulation in $8$ seconds.

Fig.~\ref{fig:trajectory} shows the states $x_{i}$. It can be observed that the trajectories are synchronized to the common trajectory. Actually, the synchronization error $\delta (t)$ is smaller than the theoretically guaranteed tolerance $\epsilon$ after $t\approx4.5$[s] as shown in  Fig.~\ref{fig:delta}. The synchronization error reaches 0.0032 in the simulation time, whereas $\epsilon$ in Theorem~\ref{Thm1} is equal to 0.0637. Fig.~\ref{fig:acceses} indicates the accesses of the agents within the interval from $t=5$[s] to $t=8$[s]. It can be seen that the sequence of the access times of each agent does not have any accumulation points. As shown in Table \ref{table:cloudaccess}, the average access intervals and the minimal access intervals are much larger than the lower bound $\tau_{i}^{\star}$ for all the agents.
\begin{table}[htb]
\begin{center}
\caption{Cloud access}
\label{table:cloudaccess}
\begin{tabular}{c c c c}
\hline
\mbox{\textbf{}}& \textbf{Access} & \textbf{Minimum interval}& \textbf{Average interval}\\\hline
\mbox{\scalebox{1.17}{1}} & $68$ & $0.0182$ & $0.1164$\\ 
\mbox{\scalebox{1.17}{2}}& $69$ & $0.0161$ & $0.1153$\\ 
\mbox{\scalebox{1.17}{3}}& $68$ & $0.0302$ & $0.1185$ \\ 
\mbox{\scalebox{1.17}{4}} & $67$& $0.0329$ & $0.1182$\\ \hline
\end{tabular}
\end{center}
\end{table}%

\section{Concluding Remarks}
In this paper, we have proposed a self-triggered controller for the bounded synchronization problem of the \emph{high-order} LTI multi-agent system based on the asynchronous information exchange through a cloud repository. We have designed an access rule by tightly evaluating the bound on matrix exponentials. We have also proved that the proposed method is feasible in the sense that the closed-loop system does not exhibit any Zeno behavior. As a future work, it remains to extend the results of this paper to the case where the system matrices $(A,B)$ have parametric uncertainties.

\section*{Acknowledgements}
This work is supported by JST-SPRING JPMJSP2101.

\bibliographystyle{plain}       

\appendix
\end{document}